# Exploring Strategies for Personalized Radiation Therapy Part IV: An Interaction-Picture Approach to Quantify the Abscopal Effect

Hao Peng, Laurentiu Pop, Kai Jiang, Faya Zhang, Debabrata Saha, Raquibul Hannan, Robert Timmerman

Department of Radiation Oncology, University of Texas Southwestern Medical Center, Dallas, TX 75390, USA.

**Corresponding Author:**

Hao Peng, PhD, Email: Hao.Peng@UTSouthwestern.edu

**Running title:** Abscopal effect and interaction picture

The authors have no conflicts to disclose.

Data sharing statement: The data that support the findings of this study are available from the corresponding authors upon reasonable request.

**Abstract:**

We revisit the controversial "abscopal" effect in the context of Personalized Ultra-Fractionated Stereotactic Adaptive Radiotherapy (PULSAR). By allowing long interval between fractions, PULSAR may enhance systemic immune activation and increase the likelihood of abscopal responses compared with conventional daily fractionation. To quantify treatment-induced effects, we introduce an interaction-picture transformation adapted from quantum mechanics, which separates intrinsic tumor growth from radiation- and immune-mediated perturbations. In this preliminary study, we tested this method to two preclinical bilateral tumor models (4T1 and MC38). Our model provides a quantitative measure of the interaction strength between primary and secondary tumors at the individual level, capturing dynamics over time rather than relying solely on cohort averages. This approach frames the abscopal effect as a continuous, stochastic phenomenon rather than a binary response. The framework is flexible for future studies, particularly in concurrent radiation and immunotherapy with PULSAR, where different radiation doses and fractionation schedules can be compared, and immune checkpoint inhibitors (ICIs) can be incorporated to further enhance systemic anti-tumor immunity. The framework can also help us make cross-study comparison of abscopal effects and standardizes the reporting of abscopal magnitude beyond simple statistical significance.

## 1. Introduction

The abscopal effect, first described by R.H. Mole in 1953, refers to radiation-induced tumor regression at sites distant from the irradiated field [1]. While the primary mechanism of radiotherapy involves direct DNA damage and free radical formation through radiolysis—resulting in local tumor control, growing evidence supports the existence of radiation-induced abscopal effects [2-8]. These effects occur in non-irradiated cells located near or far from the treated site and can be categorized by a spatial scale into bystander, abscopal, and cohort effects [2]. The bystander effect typically occurs over short cellular distances (millimeters), whereas the abscopal effect manifests at distant metastatic sites, often several centimeters away.

Despite increasing interest, the abscopal effect remains poorly understood. A systematic review identified 46 reported cases between 1969 and 2014 [3]. These cases span multiple malignancies, including melanoma, leukemia, and renal cell carcinoma, particularly among highly immunogenic tumors [4-7], suggesting that radiotherapy can modulate the tumor microenvironment and elicit systemic effects beyond local tumor control. Notably, abscopal responses may not always be beneficial. In one recent study, radiation can induce amphiregulin expression in tumor cells, which promotes distant metastasis by shifting EGFR-expressing myeloid cells toward an immunosuppressive phenotype with reduced phagocytic activity. This phenomenon, referred to as a "*badscopal*" effect [8], highlights the potential for adverse systemic consequences rather than therapeutic antitumor responses. Studies have also shown that radiotherapy alone may not be insufficient to generate durable local and abscopal responses [7, 9, 10]. The advent of immunotherapy has helped overcome tumor-induced immunosuppression, making the combination of radiotherapy and immunotherapy a promising strategy to enhance systemic antitumor effects. One study reported that the combination of immune checkpoint inhibitors (ICI) and radiation significantly prolonged progression-free survival (PFS) and overall survival (OS) in patients with lung cancer compared with ICI alone [9]. Another study demonstrated that patients who received radiation prior to ICI initiation had a significantly higher lung lesion response rate than those who received RT after starting, with a trend toward improved OS and PFS [10].

Its rarity, difficulty in quantification, heterogeneity across tumor types, and complex immunologic mechanisms make investigation of the abscopal effect challenging. Several representative studies are briefly summarized below. The first line of research focuses on identifying candidate biomarkers and clinical correlates that may predict which patients will experience an abscopal effect; however, no robust or universally validated predictors have yet emerged [12–16]. A pooled analysis of three phase I/II trials combining immunotherapy and radiotherapy demonstrated that higher post-treatment absolute lymphocyte count (ALC) was associated with increased likelihood of abscopal responses [14]. Preclinical studies integrating radiomics features with peripheral blood markers showed promising predictive performance in a single murine model [15]. In addition, mathematical modeling suggests that T-cell trafficking and immune activation alone are insufficient to fully explain the occurrence of abscopal effects, indicating that additional systemic and microenvironmental factors are involved [16]. The second line of investigation focuses on the relationship between radiation dose and the induction of the abscopal effect [5–7]. Evidence suggests that conventional low-dose radiation (<2 Gy per fraction) is insufficient to trigger systemic anti-tumor responses. In contrast, higher doses (>8 Gy per fraction) and selected hypofractionated regimens appear more likely to stimulate immune-mediated effects. However, excessively high doses (>20 Gy per fraction) may attenuate this benefit, potentially due to vascular damage and reduced immune cell infiltration within the tumor microenvironment [5–7]. Furthermore, one meta-analysis reported that a biologically effective dose (BED) of approximately 60 Gy is associated with an estimated 50% probability of observing an abscopal response [11].

Our institute is investigating Personalized Ultra-Fractionated Stereotactic Adaptive Radiotherapy (PULSAR), a special form of Stereotactic Body Radiation Therapy (SBRT) characterized by prolonged intervals between high-dose radiation pulses [17–21]. By temporally spacing treatment, PULSAR can better align with the kinetics of anti-tumor immune activation, allowing tumor-specific T cells to recover and clonally expand between fractions, which may increase its potential to induce the abscopal effect compared with conventional daily fractionation. In contrast, continuous low-dose schedules can result in cumulative lymphocyte depletion and chronic immunosuppressive remodeling of the tumor microenvironment, thereby limiting systemic immune responses. Moreover, the adaptive timing of PULSAR enables strategic coordination with immune checkpoint inhibitors, facilitating iterative immune re-stimulation analogous to a booster vaccination approach. Collectively, these features provide a strong rationale to investigate whether PULSAR can enhance systemic immune amplification and improve the likelihood of clinically meaningful abscopal responses.

A central challenge in studying the abscopal effect is disentangling treatment-induced responses from intrinsic tumor growth (**Figure 1**). To address this, we adopt an interaction-picture transformation, a mathematical framework adapted from quantum mechanics [22]. In this formulation, the baseline growth dynamics of each tumor are factored out, yielding a transformed tumor state that captures only the additional perturbations attributable to radiation or systemic (abscopal) interactions. This approach enables direct quantification of treatment-induced deviations relative to natural tumor progression and facilitates comparison between the rapid, localized effects of radiation and the more gradual emergence of distal responses. By isolating both the magnitude and temporal evolution of local and systemic effects, the framework provides a clearer view of how radiation modulates tumor–tumor interactions and the conditions under which abscopal responses arise. In this proof-of-concept study, we applied the framework to two widely used preclinical tumor models, 4T1 and MC38, each incorporating primary and secondary tumors to permit controlled measurement of both local radiation effects and distant responses. As a pilot investigation, our objective is methodological development and validation of feasibility, rather than definitive mechanistic characterization of biological processes underlying the abscopal effect.

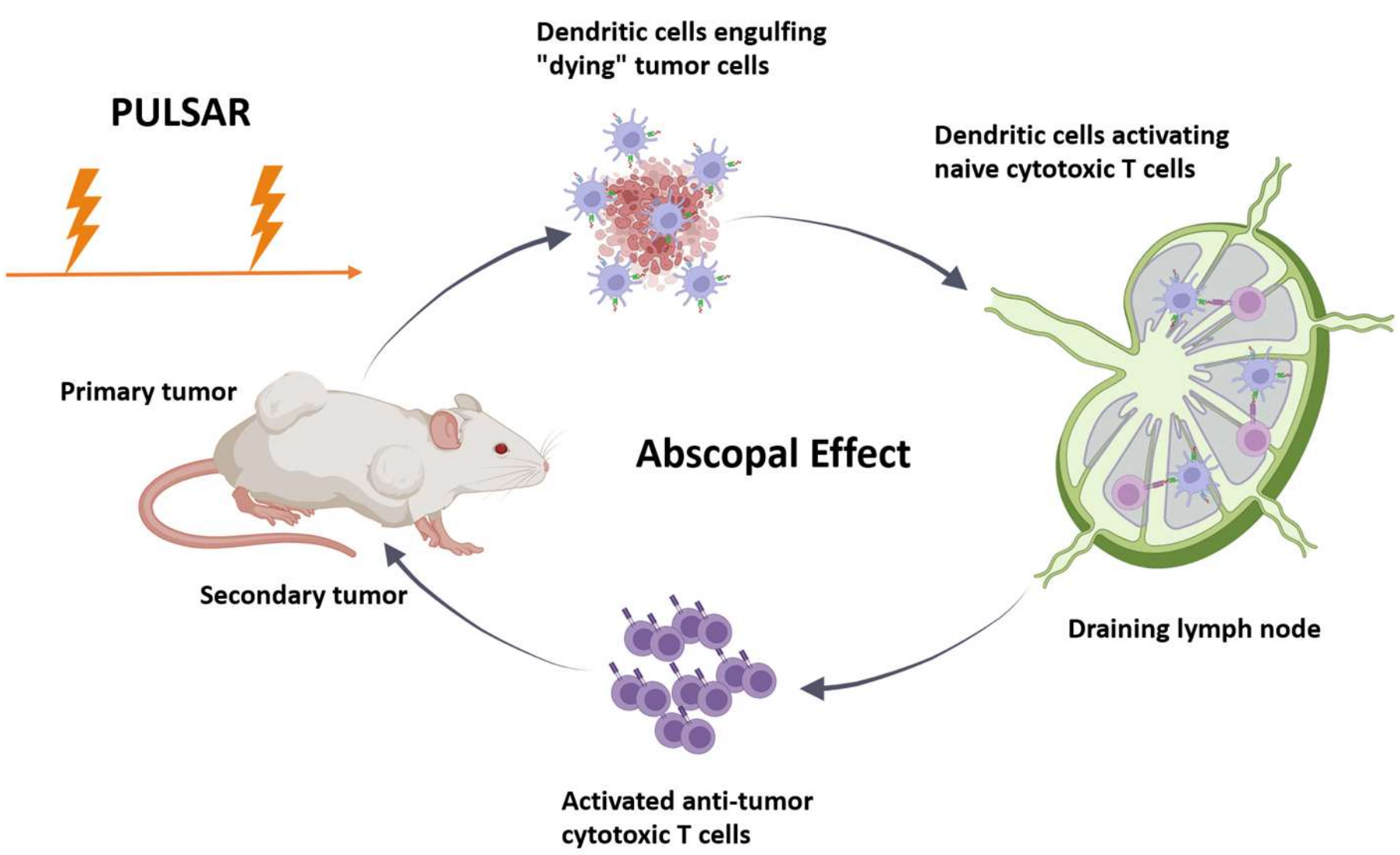

**Fig. 1.** Abscopal effect and PULSAR. Radiation dose and timing in PULSAR may impact the kinetics of anti-tumor immune activation and increase its potential to induce the abscopal effect compared with conventional daily fractionation.

## 2. Methods

We apply the interaction picture and perturbation, a concept from quantum mechanics to model tumor growth under treatment. Without treatment, tumor growth follows an exponential or a logistic model representing the intrinsic dynamics: autonomous growth driven solely by internal tumor properties, independent external factors. Treatment is viewed as a time-dependent perturbation to this system, representing radiation effects such as cell killing or immune activation that vary with dose, timing, and cyclic immunotherapy if used. This leads to a perturbed ordinary differential equation (ODE) analogous to the perturbation theory in quantum physics. Using the interaction picture allows us to: 1) decouple intrinsic tumor growth from treatment effects by solving the baseline dynamics first; 2) track how treatment modifies tumor evolution as a perturbation on that baseline; and 3) efficiently approximate dynamics when radiation or drug effects are slowly varying. We want to solve a differential equation of the form where $x$ represents the tumor volume, and the term A has two parts:

$$\frac{dx}{dt} = Ax(t) = (A_0 + B)x(t) \quad (1)$$

where $A_0$ reflects the intrinsic growth dynamics which captures the natural evolution of the system without any external intervention or interaction effects. $B$ is the interaction or treatment matrix which represents additional effects beyond the intrinsic dynamics, revealing the existence of potential abscopal effects. Such decomposition helps isolate and study treatment or interaction dynamics separately from the natural system evolution. In other words, it moves the system into the interaction picture, removing the evolution due to $A_0$.

### 2.1 Treatment effect

As the first step to test the approach, we generated tumor growth dynamics using a logistic growth model with and without treatment and developed a model to decompose the total growth rate into intrinsic and treatment-induced components. The untreated tumor size $V(t)$ was simulated according to the logistic growth equation. Numerical integration was performed using an explicit Euler method with step size $\Delta t$. The intrinsic growth rate is defined as $a(t)$ and treatment is modeled as a localized pulse centered at time $t_{peak}$ with Gaussian temporal profile $b(t)$. Solutions for the untreated $V(t)$ and treated $x(t)$ can be generated numerically with initial value set to be one:

$$V(t) = \exp\left(\int_0^t a(s)ds\right) \qquad \text{x(t)} = \exp\left(\int_0^t [a(s) + b(s)]ds\right) \tag{2}$$

Given the synthetized data, we first estimate the intrinsic rate $\hat{a}(t)$ from untreated data *V(t)* by fitting and differentiating to get the "interaction picture" tumor state which isolates treatment-induced dynamics by factoring out the intrinsic evolution. The treatment effect $\hat{b}(t)$ can also be estimated.

$$\hat{a}(t) = \frac{1}{V(t)}\frac{dV(t)}{dt} \qquad x_I(t) = \frac{x(t)}{V(t)} \quad \hat{b}(t) = \frac{1}{X_I(t)}\frac{dx_I}{dt} \tag{3}$$

**2.2 Abscopal effect**

We modeled the abscopal effect as a coupled linear system and applied the interaction picture formalism to separate intrinsic dynamics from coupling effects. Several assumptions are made to simplify the problem. First, it treats each tumor (primary or secondary) as having separable, intrinsic dynamics—its own growth and treatment response—while the abscopal effect is added on top as a perturbation that does not alter the baseline form. The interaction is written as a simple cross-term, usually linear, even though biologically the immune-mediated effect is nonlinear, involving thresholds, saturation, or delays. By representing the coupling with an off-diagonal term, the model also assumes direct, pairwise influence between tumors, whereas the connection is mediated through more complicated immune processes.

The primary-secondary tumor population vector $x(t) = [X_P(t), X_S(t)]$ evolved according to Eqs. (1) and (4) where:

$$\frac{d}{dt}\begin{bmatrix} X_P \\ X_S \end{bmatrix} = \begin{bmatrix} r_P & 0 \\ 0 & r_S \end{bmatrix}\begin{bmatrix} X_P \\ X_S \end{bmatrix} + \begin{bmatrix} \alpha & 0 \\ \beta & 0 \end{bmatrix}\begin{bmatrix} X_P \\ X_S \end{bmatrix} \tag{4}$$

The diagonal matrix $A_0$ represents uncoupled growth terms (same intrinsic growth rate for both primary and secondary tumors). The matrix *B* encodes the treatment effect and coupling, where $\alpha$ models radiation-induced tumor killing and $\beta$ models abscopal-mediated tumor killing (one way interaction from the primary to the secondary). Define the interaction-frame state $X_I(t)$ and the interaction-frame operator $B_I(t)$. Their product $B_I(t)X_I(t)$ gives the effect of perturbations (time dependent) excluding baseline exponential tumor growth of both primary and secondary tumors. Letting $r_S$=$r_P$ makes $A_0$ and B commute and the exponential terms cancel in Eq. (5). However, the model can easily extend to model primary and secondary tumors of different growth rates.

$$X_I(t) = e^{-A_0 t}X(t) \qquad \frac{dX_I(t)}{dt} = e^{-A_0 t}Be^{A_0 t}X_I(t) = B_I(t)X_I(t) \tag{5}$$

Once the tumor trajectories under treatment are available, the model numerically computes derivatives to back-calculating $\alpha$ and $\beta$. To smooth the tumor curves and minimize fluctuation, we test exponential fitting, linear, and cubic splines. It then constructs an interaction picture by factoring out baseline exponential growth.

$$\dot{X}_P = r_P X_P + \alpha X_P \qquad \alpha(t) = \frac{\dot{X}_P(t) - r_P X_P(t)}{X_P(t)} \cong \hat{\alpha} - r_P \tag{6}$$

$$\dot{X}_S = r_S X_S + \beta X_P \qquad \beta(t) = \frac{\dot{X}_S(t) - r_S X_S(t)}{X_P(t)} \cong (\hat{\beta} - r_S)\frac{\hat{X}_S(t)}{\hat{X}_P(t)} \tag{7}$$

where $\hat{\alpha}$ and $\hat{\beta}$ are estimated from exponential fitting. $\alpha(t)$ in the matrix $B$ will be constant, while $\beta(t)$ in the matrix $B$ depends on the volume of both tumors (i.e., entangled) and time. Two subtleties are worth emphasizing. First, estimation of $\beta(t)$ will be highly sensitive to fitting error, but $B_I(t)X_I(t)$ avoids this by taking into account volume by multiplying $X_I(t)$ in Eq. (5). Second, the abscopal effect is modeled as a one-way interaction from the primary to the secondary tumor, only the primary-tumor component in $X_I(t)$ contributes to the coupling (i.e., independent of secondary tumor).

### 2.3 Animal experiments

Female NCI C57BL/ (556) and NCI BALB/C (555) 7-8 weeks old were purchased from Charles Rive Laboratories (Wilmington, MA). All mice were maintained under specific pathogen-free conditions within the facilities of the Animal Resource Center at the University of Texas Southwestern Medical Center (UTSW). Animal protocol was approved by the Institutional Animal Care and Use Committee at UTSW. Murine colon cancer cell line MC-38 was purchased from Kerafast, Inc. (Shirley, MA) and murine breast cancer line 4T1 was purchased from ATCC (Manassas, VA). MC38 and 4T1 cell lines were cultured in 10% fetal calf serum (Gibco, Waltham, MA) Dulbecco's Modified Eagle's Medium (DMEM) and Roswell Park Memorial Institute (RPMI) 1640, respectively. Cell cultures were maintained at 37 °C in a humidified atmosphere containing 5% CO2. The cell lines were regularly verified for mycoplasma using the Universal Mycoplasma Detection Kit from ATCC.

**Table 1. Summary of radiation dose and timing for animal studies.**

| *4T1*<br>*Group* | *Right tumor RT*<br>*dose/schedule* | *Left tumor RT*<br>*dose/schedule* |
|---|---|---|
| **1** | 0 Gy, 0 Gy | N |
| **2** | 16 Gy, 8 Gy<br>D0, D1 | N |
| **3** | 16 Gy, 8 Gy<br>D0, D10 | N |

| *MC38*<br>*Group* | *Right tumor RT*<br>*dose/schedule* | *Left tumor RT*<br>*dose/schedule* |
|---|---|---|
| **4** | 0 Gy, 0 Gy | N |
| **5** | 8 Gy, 4 Gy<br>D0, D1 | N |
| **6** | 8 Gy, 4 Gy<br>D0, D10 | N |
| **7** | 12 Gy / D0 | N |
| **8** | 8 Gy / D0 | 4 Gy / D0 |

The summary of experimental studies is provided in **Table 1**. For 4T1 model, NCI BALB/C (555) mice were inoculated s.c. in the right hind leg (primary tumor) with 0.5 x $10^6$ cells and in the left hind leg (secondary tumor) with 0.02 x $10^6$ cells. For MC-38 model, NCI C57BL/ (556) mice were inoculated subcutaneously (s.c.) in the right hind leg (primary tumor) with 0.5 x $10^6$ cells and in the left hind leg (secondary tumor) with 0.25 x $10^6$ cells. Once the tumor size reached about 7-8 mm in diameter, primary tumors were locally irradiated with different radiation schedules designed for both tumor models as shown in Table 1 using an X-RAD 320 irradiator (Precision X-Ray). In only one experimental arm group for MC-38 model, the secondary tumor was irradiated, too (**Table 1**). The irradiation was delivered with the gantry directly facing the tumor, and the irradiation beam was controlled by a collimator with an appropriate opening matching the tumor's diameter. Standard dosimetry was performed to calculate the delivered dose to the center of the tumor. Tumor sizes were measured by a caliper three times a week until the tumor reached 2.0 cm in diameter, which was the endpoint of the experiment per protocol. Tumor volumes (TVs) were calculated using the following formula: TV ($mm^3$) = length (mm) × width (mm) × width (mm) / 2.

## 3. Results

### 3.1 Two simplified examples

The core concepts of the proposed modeling framework are illustrated in two examples using a synthetically generated dataset. These toy models are not intended to capture the full biological complexity of tumor dynamics, but rather to demonstrate the underlying mathematical and computational principles. In **Figure 2**, we show the true and estimated treatment effects, tumor trajectories with and without treatment, and the corresponding interaction picture. **Figs. 2A** and **2B** represent two different treatment scenarios, peaking at different times (day 2 and day 10), respectively). By comparing the $x_I(t)$ defined in Eq. (3), the framework enables quantification of treatment effects. $x_I(t)$ =1 indicates that the interaction picture perfectly mirrors intrinsic dynamics, implying no detectable treatment or interaction effect. $x_I(t)$ <1 indicates suppression of tumor growth relative to the natural trajectory, reflecting negative effects such as treatment-induced inhibition.

In **Figure 3**, primary and secondary tumor trajectories are simulated, each with their own growth constant and abscopal effect. The interaction picture, $B\times X$, and $\beta$ are shown. The primary tumor serves as the driving term, while the secondary tumor behaves as a system under drive. **Fig. 3A** shows the raw measured trajectories for both tumors, representing intrinsic baseline behavior. **Fig. 3B** displays the secondary tumor in the interaction picture, which removes the intrinsic growth component. Deviation of $x_I(t)$ from 1 reveals suppression relative to baseline due to abscopal effect. **Fig. 3C** presents the derivative of the interaction picture $dx_I/dt$, capturing the instantaneous rate of change in the secondary tumor. This time-resolved view highlights periods of maximal suppression. **Fig. 3D** quantifies the abscopal effect. For $\beta$=0, it means no abscopal effect; the secondary tumor grows according to its intrinsic baseline. For $\beta$<0, the secondary tumor is suppressed relative to natural growth, indicating a negative (inhibitory) effect mediated by the primary tumor. These toy examples demonstrate that the interaction-picture framework can reliably distinguish intrinsic tumor dynamics from externally imposed effects.

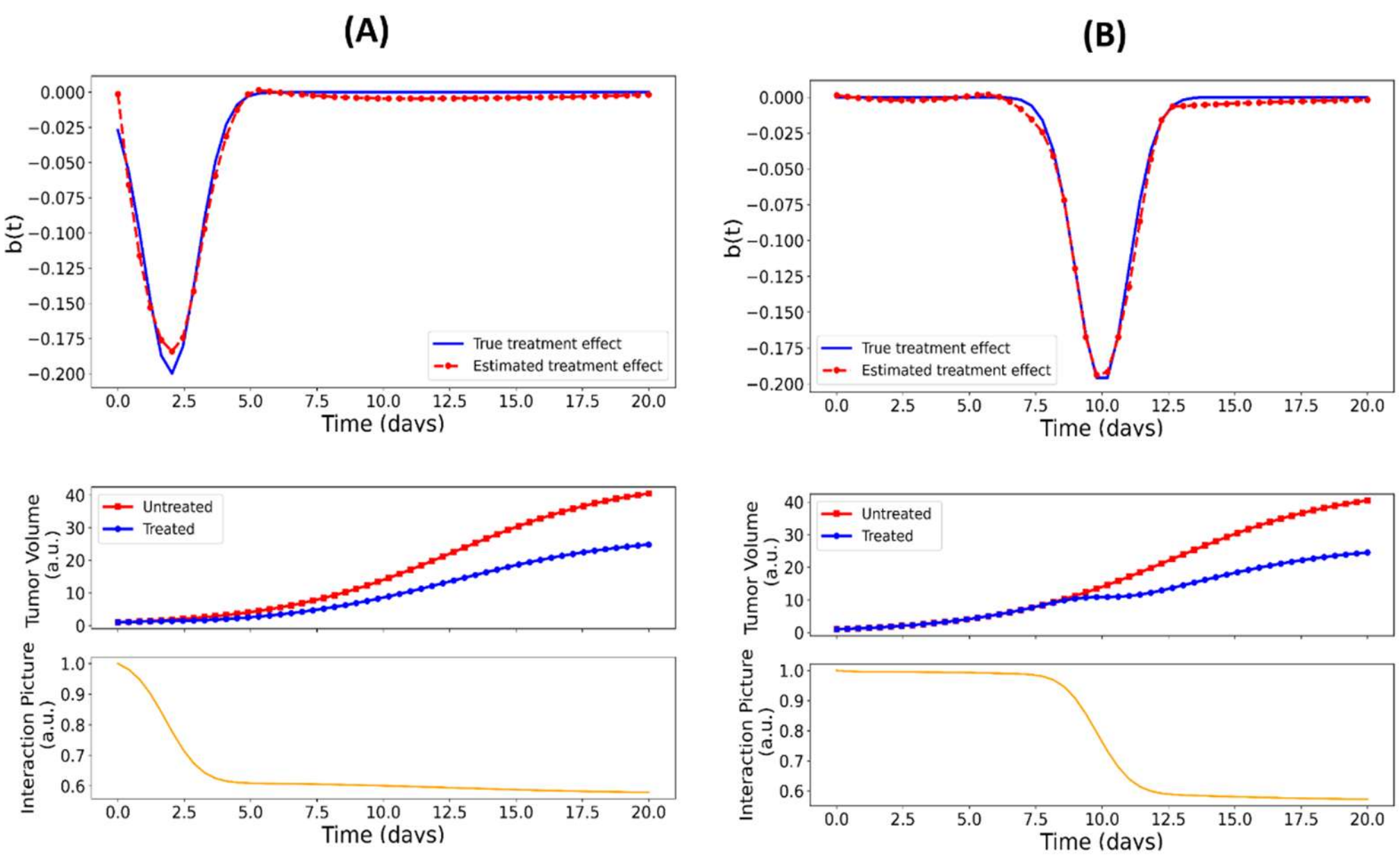

**Fig. 2.** Modeling of tumor trajectories without and with treatment (left: on day 2, right: on day 10) and the corresponding interaction picture $x_I$ defined in Eq. (3).

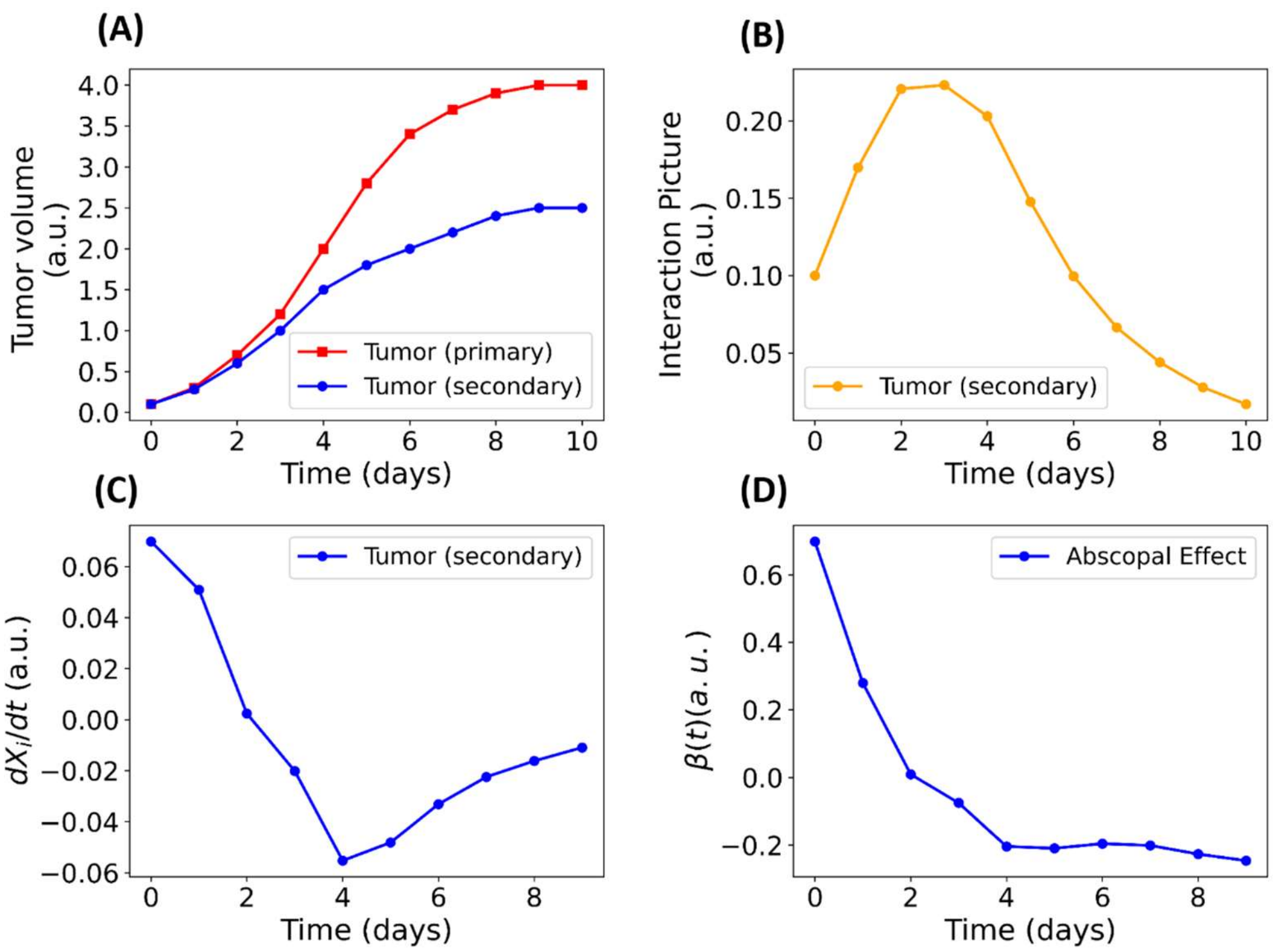

**Fig. 3.** Modeling primary and secondary tumor trajectories using synthetic data (intrinsic growth rate: 0.5 per day, unitless) and their interaction-frame representation following Eqs. (5) and (7).

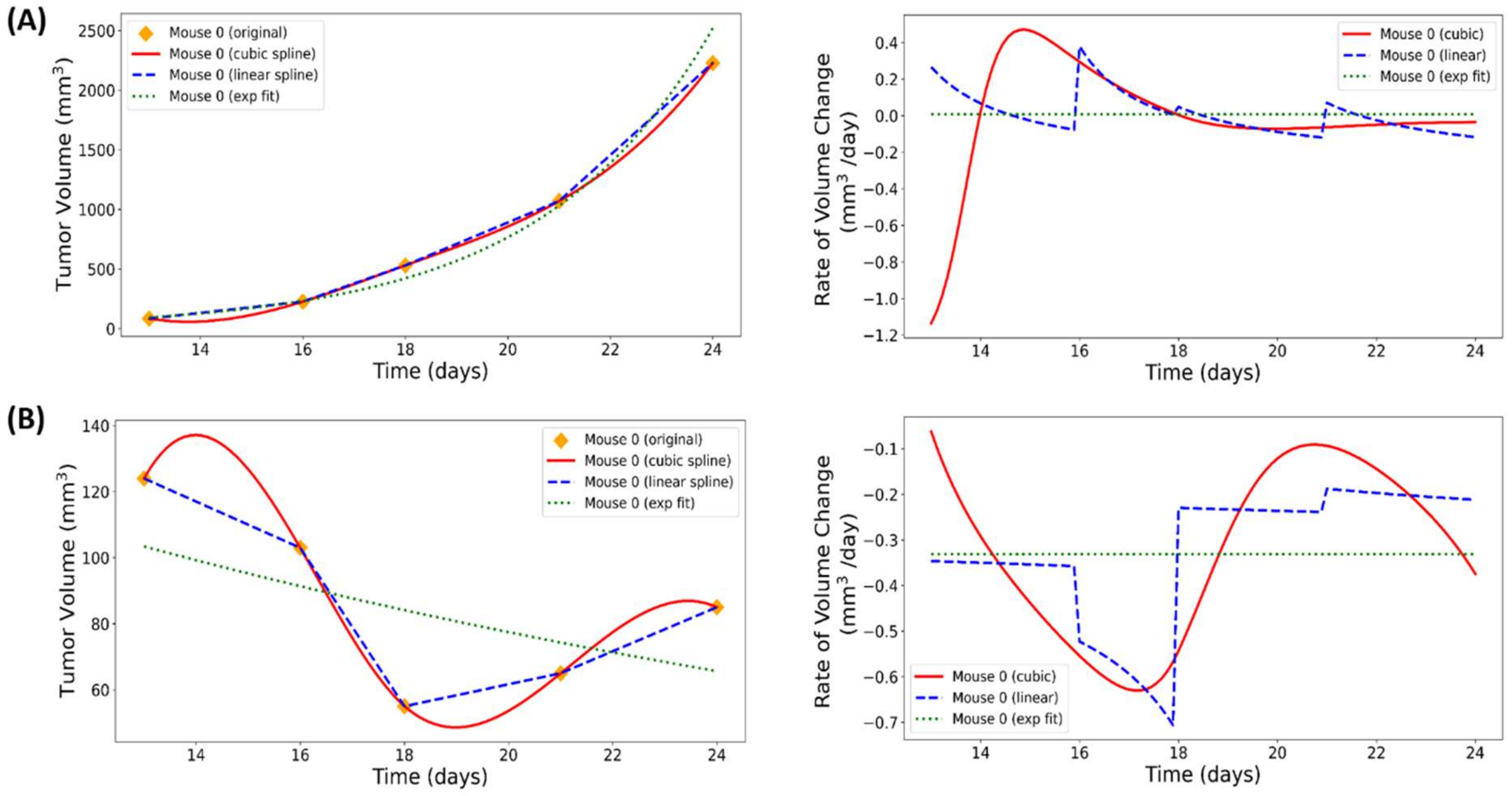

**Fig. 4.** Comparison of three fitting methods for estimating $\alpha(t)$ and $\beta(t)$ in in the matrix $B$ based upon Eqs. (6) and (7), for one primary (A) and secondary (B) tumor pair.

### 3.2 Animal experiments

Mouse tumor volume measurements are inherently noisier than the synthetic dataset due to both inter-animal variability and human measurement error. To reduce noise in individual trajectories, smoothing is required, and an exponential fit is preferred for capturing the overall trend of tumor response instead of instantaneous fluctuation as shown in **Figure 4**. Raw tumor volumes for the primary and secondary tumors are shown for three fitting approaches: cubic spline, which captures nonlinear trends while reducing noise; linear spline, a piecewise linear approximation; and exponential fit. For each method, the instantaneous rate of change reveals the difference between the observed and intrinsic growth rates, normalized by tumor size. Exponential fitting produces a constant rate of change, while two spline methods reveal transient fluctuations caused by derivative noise: even small measurement errors can produce large spikes in the derivative. Consequently, the exponential fit is preferred for highlighting the overall tumor response trend, despite its imperfect alignment with the measured volumes.

For each mouse, we analyzed the paired trajectories of the primary and secondary tumors, as illustrated in **Fig. 5**. Detailed results for all experimental groups are provided in the Supplementary Materials. For the 4T1 model, inter-animal variability was minimal, and all seven mice were included in the analysis. In contrast, MC38 tumors exhibited greater variability; therefore, we selected four mice whose growth trajectories aligned most closely with the cohort-averaged growth rate for the abscopal analysis. To establish a reasonable baseline for intrinsic tumor growth, we estimated the maximum growth rates, yielding 0.29 $mm^3$/day for MC38 and 0.17 $mm^3$/day for 4T1 tumors. Although using the maximum growth rate may introduce a slight upward bias in estimating the treatment effect and reduce variability among mice within each group, we believe this approach provides a reasonable reference for isolating treatment-induced effects and can be improved as more mice per group become available.

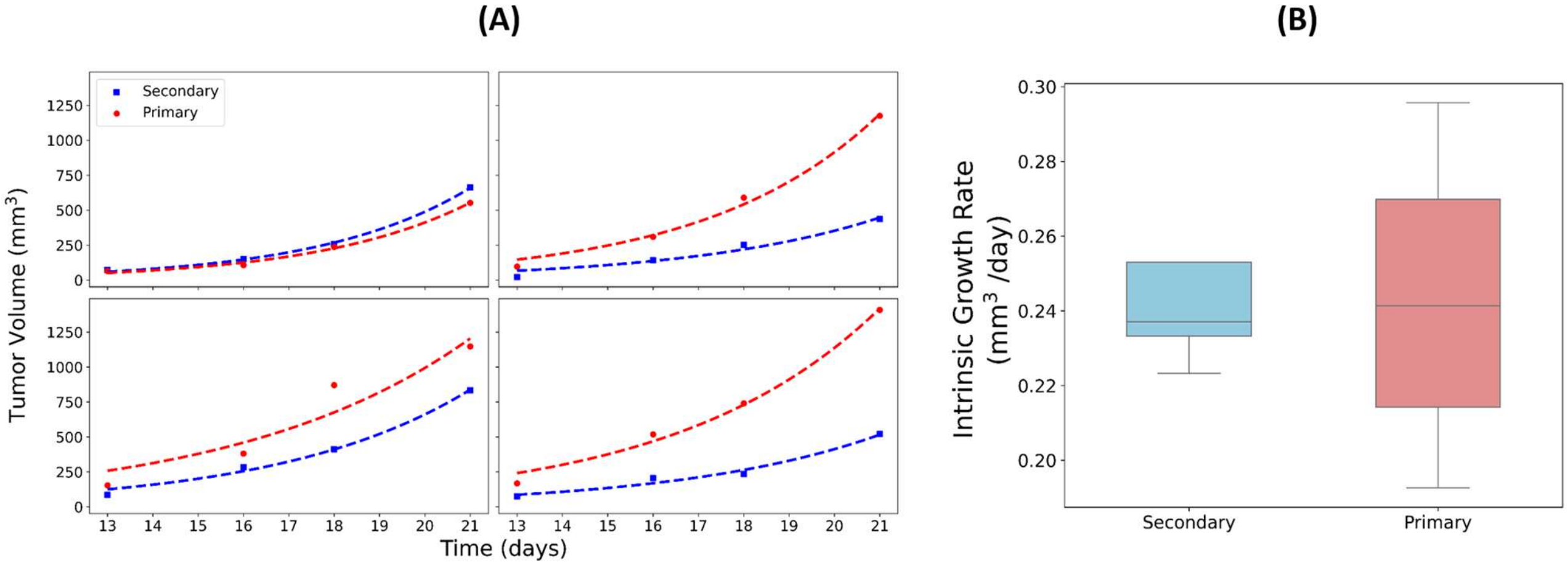

**Fig. 5.** Primary and secondary tumor trajectories analyzed after fitting intrinsic growth using the MC38 model (group 4), yielding a mean trajectory with derived $r_S$ and $r_P$ in Eq. (4) as the cohort average.

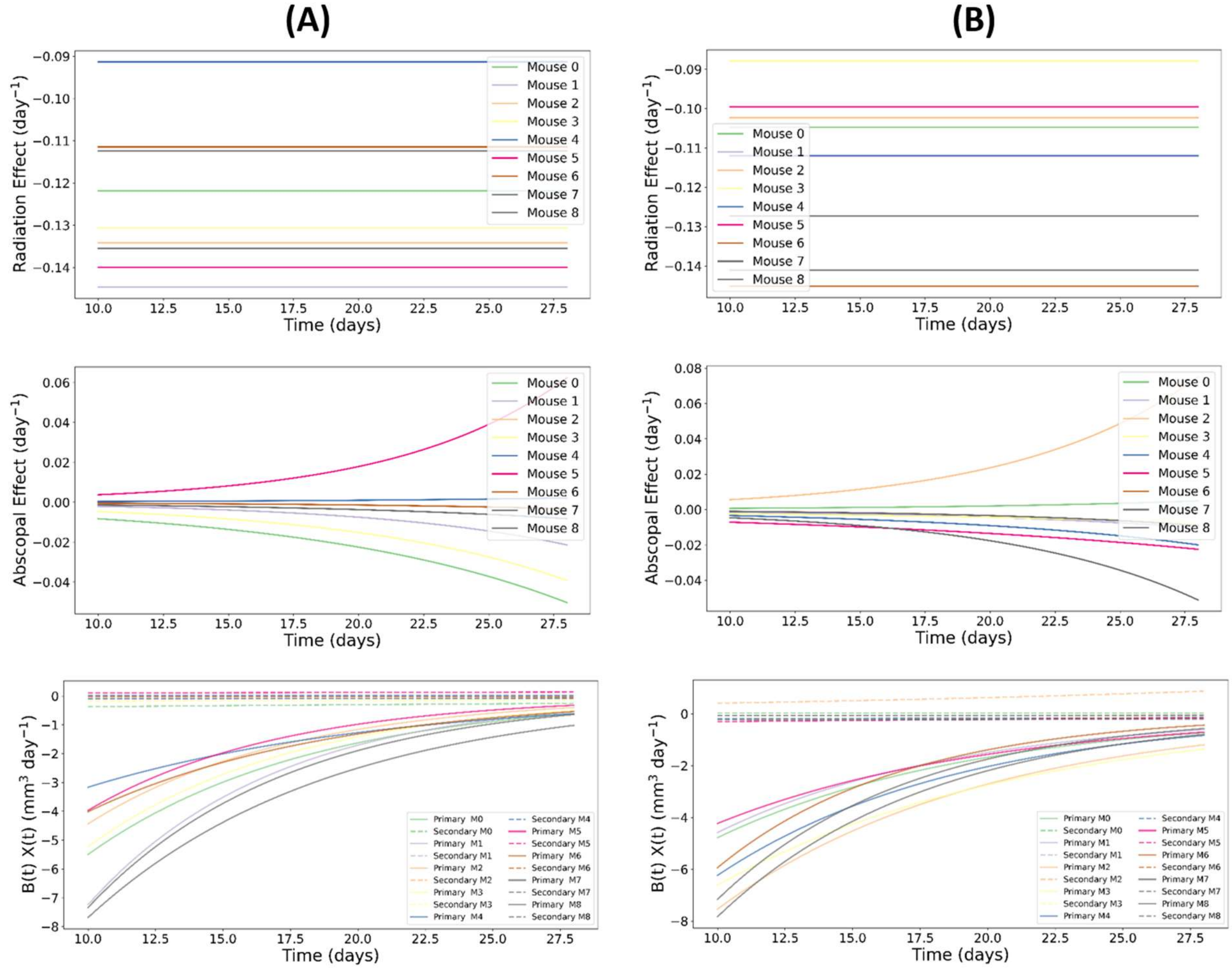

**Fig. 6.** Results of 4T1 model: A (group 2) and B (group 3). Top: Radiation-induced effect, represented by $\alpha(t)$. Middle: Abscopal effect, represented by $\beta(t)$. Bottom: Combined perturbation term, $B_I(t)X_I(t)$ (see Eq. (5) for the definition).

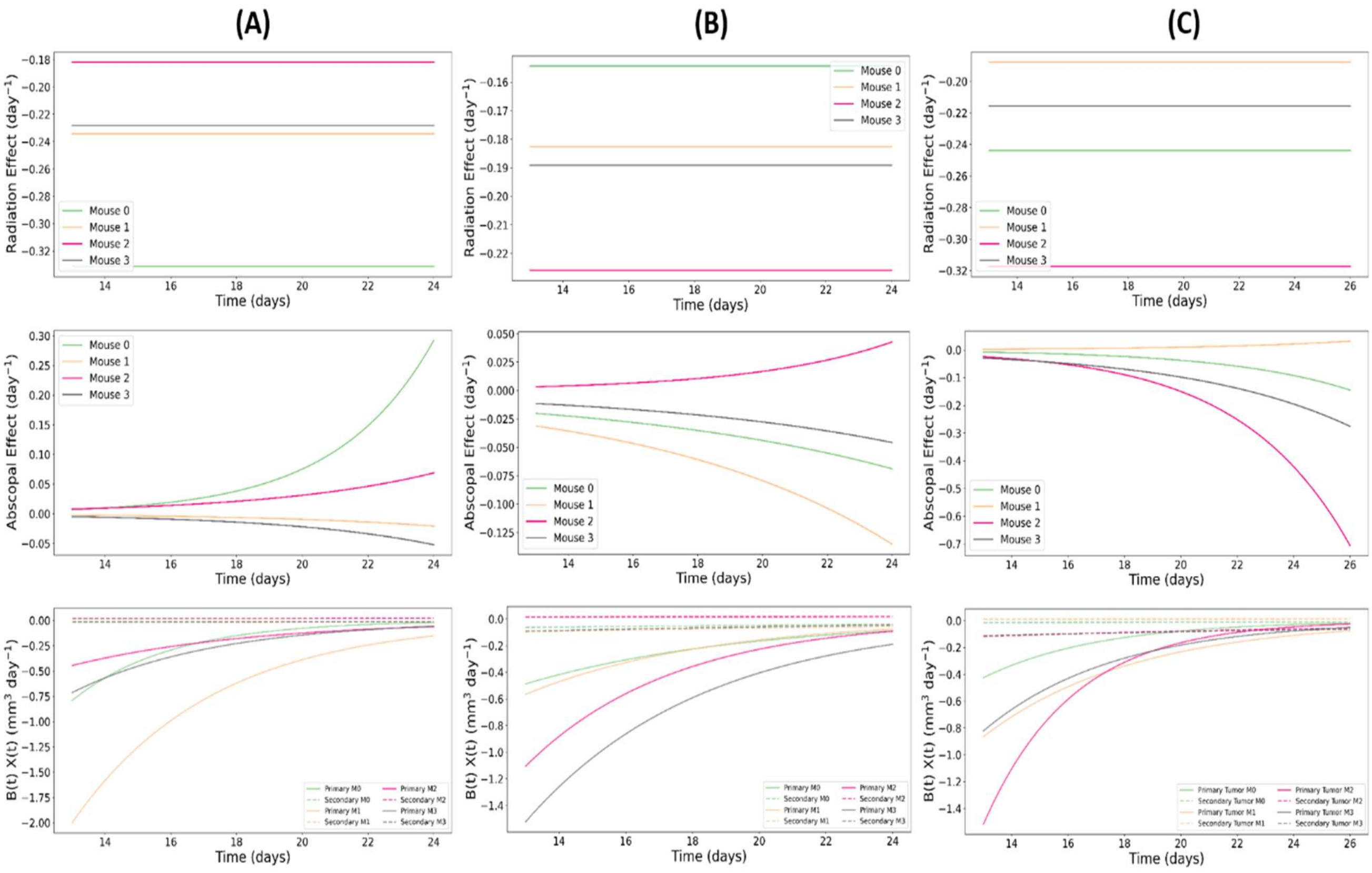

**Fig. 7.** Results of MC38 model, A (group 5), B (group 6) and C (group 7), presented in the same format as Fig. 6.

The results are summarized in **Fig. 6** for the 4T1 model and **Fig. 7** for the MC38 model. The top panels show the radiation-induced killing term $\alpha(t)$, the middle panels show the abscopal coupling term $\beta(t)$, and the bottom panels display the combined perturbation term $B_I(t)X_I(t)$. The radiation term $\alpha(t)$ is consistently negative, reflecting direct tumor cell killing by irradiation. Owing to exponential fitting of tumor trajectories, both the numerator and denominator in the estimation of $\alpha(t)$ are exponential functions, yielding a constant ratio (see Eqs. 5 to 7). Small variations in $\alpha(t)$ (ranging from −0.15 to −0.09) arise primarily from inter-animal variability. The abscopal coupling term $\beta(t)$ is negative, except in one mouse, indicating a suppressive effect of the irradiated primary tumor on secondary tumor growth. Its magnitude is weaker than that of the direct radiation effect, typically ranging between −0.05 and 0.05, and exhibits mild temporal variability. The combined perturbation term $B_I(t)X_I(t)$ captures the net biological effect of radiation and immune-mediated coupling, after removing intrinsic exponential tumor growth. For the primary tumor, this term exhibits a strong negative excursion immediately following irradiation (16 Gy on day 0 and 8 Gy on day 1 for Group 2), followed by gradual relaxation toward zero, as radiation-induced damage dissipates. A similar temporal pattern is observed for the fractionated regimen with a 10-day separation (16 Gy on day 0 and 8 Gy on day 10 for Group 3). These results indicate that radiation delivers an acute inhibitory impulse that produces a lasting but progressively weakening suppression of tumor burden, with the absolute killing rate decaying from approximately −4 to −8 volume units per day toward zero. In contrast, the abscopal contribution to the secondary tumor remains much weaker and stable over time, typically about 0 to -0.5 volume units per day.

Note that in linear exponential growth models, the absolute rate of tumor change is proportional to the current tumor volume, explaining the observed decay in killing as tumor size decreases. Although radiation is delivered as single pulse (e.g., day 0), its biological consequences unfold over extended timescales through delayed cell death and immune activation, producing an effective exponential decay. Thus, α represents the persistent "ringing" of biological response rather than the instantaneous radiation "strike." Overall, β remains much smaller than α, consistent with the biological expectation that direct radiation induces strong acute tumor killing, whereas the abscopal effect exerts a weaker but sustained immunomodulatory influence. The interaction-picture formulation therefore isolates the true net biological action of radiation and immune coupling, uncontaminated by intrinsic tumor growth.

For MC38, similar trends are observed across the three treatment groups (**Fig. 7**). For the primary tumor, the combined perturbation term starts out strongly negative, indicating substantial tumor cell killing immediately following irradiation. This behavior is evident after the first pulse (8 Gy on day 0 and 4 Gy on day 1 for Group 5), the fractionated schedule separated by 10 days (8 Gy on day 0 and 4 Gy on day 10 for Group 6), and the single-fraction treatment on day 0 (Group 7). Consistent with prior biological observations, MC38 tumors are more radiosensitive than 4T1, which is reflected in the more negative values of $\alpha(t)$ compared with those observed for 4T1. For Group 7, a transiently elevated $\beta(t)$ reaching values as low as −0.7 is observed, which we think is due to derivative noise during numerical differentiation and the division by exceedingly small primary tumor volumes in Eq. (7). This sensitivity is mitigated in the combined perturbation term $B_I(t)X_I(t)$, which avoids explicit division by tumor volume and therefore yields more stable estimates. It shows consistent trends across all MC38 groups, with the primary tumor exhibiting a strong inhibitory response (approximately −1.4 to 2.0 volume units per day), while the secondary tumor response remains weaker, typically ranging between 0 and −0.5 volume units per day.

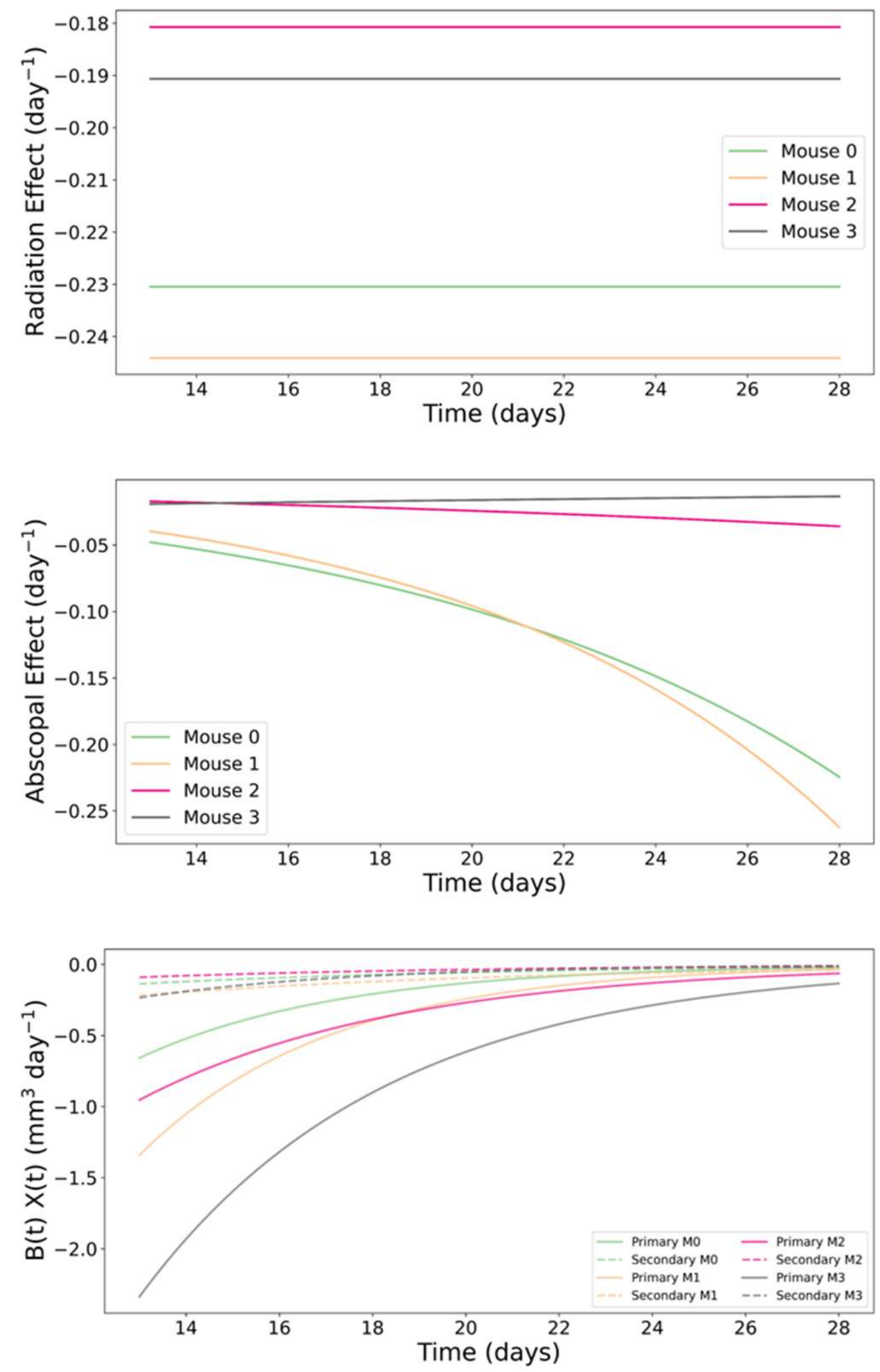

**Fig. 8.** Results of MC38 model (group 8) are presented in the same format as Fig. 6. Both primary and secondary tumors received radiation, 8 Gy and 4 Gy, respectively.

For group 8 where the secondary tumor additionally receives a 4 Gy dose, the observed trajectory reflects a combination of direct radiation killing and abscopal effect. We estimate how the abscopal effect produces tumor killing equivalent to 4 Gy of direct radiation. This ballpark comparison provides an intuitive, radiation-equivalent measure of the strength of systemic tumor control, while acknowledging that direct and abscopal effects cannot be fully disentangled within the present framework. The similarity of $B_I(t)X_I(t)$ values with and without secondary tumor irradiation suggests that the effective systemic tumor suppression mediated by the abscopal response is no stronger to that achieved by a 4 Gy local dose in this setting.

## 4. Discussion

In this exploratory study, we developed a framework based on the concept of interaction picture to study the abscopal effect. The interaction picture removes baseline exponential tumor growth, isolating the coupling from the primary tumor to the secondary tumor. This allows abscopal signaling to be studied without contamination from natural tumor expansion, which otherwise complicates analysis. The interesting question is: What component of the distant tumor's dynamics can be attributed to crosstalk, independent of its natural trajectory? This is precisely what the interaction picture isolates. The term $B_I(t)X_I(t)$ represents a pure measure of interaction, independent of whether the tumor grows quickly or slowly. It quantifies the strength of tumor crosstalk without being confounded by baseline growth rates.

Rather than relying solely on cohort-averaged statistics, the proposed framework enables us to quantify the abscopal effect at the level of individual mice. While traditional statistics can detect whether an abscopal effect exists, our proposed model quantifies the strength of the interaction between primary and secondary tumors, accounts for dynamics over time, and produces a parameter that can be compared across conditions. In our view, tumor-immune interactions should be fundamentally stochastic rather than binary processes; accordingly, the abscopal effect should be considered a continuous, distributed biological phenomenon rather than a discrete on/off response.

The inferred abscopal effect for both 4T1 and MC38 models appears too weak to meaningfully alter the growth of secondary tumor, corresponding to a magnitude lower than direct radiation on the primary tumor (**Figs. 6 and 7**). This observation is further validated by an unconventional comparison in which a 4 Gy direct irradiation is administered to the secondary tumor, with all tumor control attributed to an "equivalent" abscopal effect in the MC38 model (**Fig. 8**). However, the above interpretation should be made with caution. First, the present analysis relies on exponential smoothing of tumor trajectories, which may obscure transient dynamics. As a result, although the initial experimental design aimed to investigate the effects of dose and timing, no strong dependence was currently observed. Second, the abscopal effect may be cumulative in nature: even modest daily contributions could accumulate over time to produce a tangible impact on tumor control. Third, tumor volume alone may be an insufficient surrogate for abscopal signaling, motivating the incorporation of additional immune biomarkers in future studies to capture systemic immune activation more comprehensively.

The radiation effect initially began at a negative value and then gradually returned toward zero, reflecting the lasting effect of radiation-induced suppression. In contrast, the abscopal effect is also negative, but much weaker and appears flat. This pattern is a direct consequence of the modeling approach. In our framework, instead of representing the radiation effect as a delta function, we smooth it by fitting exponential curves to the limited time points for both primary and secondary tumors. This simplifies data analysis and minimizes random fluctuation for estimating the rate of change in tumor volume (**Fig. 4**). To be more specific, exponential fitting reduces stochastic fluctuations in $X(t)$, suppressing short-term bursts or irregularities and yielding a cleaner, though less temporally precise, estimate of $\alpha(t)$ and $\beta(t)$. As a result, $\alpha(t)$ and $\beta(t)$ are less sensitive to noise in volume measurement; the remaining sources of variability are primarily due to model misfit (if the exponential is a poor approximation) and division by small $X_p$, which represents a mathematical instability. For the same reason, $B_I(t)X_I(t)$ is a more robust quantity to evaluate.

One may argue that a truly instantaneous treatment effect would appear as a sharp temporal spike, instead of spreading in time by our modeling approach. This is indeed a limitation of our current framework. In other words, both direct radiation killing and abscopal effects become temporally blurred and appear weaker on a per-unit-time basis. Nevertheless, we believe this smoothing is acceptable provided that the overall tumor response trends are preserved. Moreover, from a biological perspective, abscopal effects are inherently continuous and gradual, even more so than direct radiation effects, making them naturally compatible with a smoothed representation. This temporal blurring also helps explain why higher radiation doses do not necessarily produce proportionally stronger inhibition in our analysis. For example, when comparing Groups 2 and 3, as well as Groups 5 and 6, which both deliver a total dose of 12 Gy but with different fractionation schedules—either 8 Gy on day 1 followed by 4 Gy on day 10, or 12 Gy delivered over the first two days—we do not observe a clear difference in the inferred treatment effects. This arises because the smoothed signal primarily reflects long-term rather than instantaneous treatment effects.

Another limitation of the exponential smoothing approach is the short observation window. With only a limited number of time points, subtle deviations arising from either direct treatment or abscopal effects are masked. A natural next step would be to estimate time-dependent parameters $\alpha(t)$ and $\beta(t)$ without assuming an exponential model (e.g., spline fitting) or a fixed interaction matrix in Eq. (4), thereby enabling the capture of transient dynamics. However, this was not implemented in the present study due to the limited sample size. If more animals per group can be included and additional longitudinal measurements become available, preferably incorporating immune-related biomarkers rather than tumor volume alone, the numerical differentiation would become more robust. Likewise, the current model assumes that the abscopal effect evolves synchronously with the dynamics of the primary tumor. If the interaction exhibits a delayed response, the model will need to be extended by introducing a correction term to explicitly account for such temporal delays.

This preliminary study does not suffice to draw definitive statements regarding the existence or magnitude of the abscopal effect. Addressing this question will require more rigorous experimental designs and characterization of underlying immune processes. Nevertheless, the proposed framework provides a flexible and quantitative platform for future investigations, particularly in the context of concurrent radiation and immunotherapy in PULSAR, where separating local and systemic treatment effects is of key interest. For example, different radiation doses and fractionation schedules can be compared. When considering concurrent radiation and immunotherapy, immune checkpoint inhibitors (ICIs) can be included in these experimental groups to enhance systemic anti-tumor immunity. By blocking inhibitory checkpoint signals in T cells, ICIs may amplify radiation-induced immune activation, potentially increasing T cell–mediated abscopal responses. Understanding the limited magnitude of the abscopal effect observed here can help set realistic expectations for radiation therapy and inform combination strategies, such as incorporating ICIs or optimizing fractionation schedules to maximize systemic tumor control.

## 5. Conclusion

Our model provides a quantitative measure of the interaction strength between primary and secondary tumors, quantifying the abscopal effect at the level of individual mice. It tells how much the primary treated tumor influences the secondary untreated tumor over time, disentangling immune-mediated systematic effect from direct radiation killing and intrinsic tumor growth. Instead of just saying abscopal effect exists or not, this allows us to make comparison of abscopal effects between groups and standardizes the reporting of abscopal magnitude beyond simple statistical significance. Other researchers can use the same method to report a numeric interaction strength, making cross-study comparison possible.